\magnification=\magstep1
%----------------------------------------------------------------------
% Astronomy & Astrophysics-like TeX-macro
% adapted from a Springer-Verlag book macro
% by Matthias Bartelmann
% September 1993, November 1994
%----------------------------------------------------------------------
\newif\ifPostScript%
\newif\ifBoldFace%
\newif\ifReferee%
\newif\ifFrames%
%
% ... set \PostScriptfalse here if postscript files should be ignored:
%
\PostScripttrue%
\ifPostScript\input epsf.tex\fi%
%
% ... set \BoldFacetrue here if bold-faced headlines requested:
%
\BoldFacefalse
%
% ... set \Refereetrue here if referee layout requested:
%
\Refereefalse
%
% ... set \Framestrue here if frames around figures requested:
%
\Framesfalse
%
% ... set \englishfalse here for german headings:
%
\newif\ifenglish\englishtrue%
%
% ... some general definitions:
%
%\magnification = \magstep1%
\ifReferee%
    \font\headfont = cmr10 at 10truept%
\else%
    \font\headfont = cmti10 at 10truept%
\fi%
\def\eventitle{}%
\def\oddtitle{}%
\ifReferee\relax\else%
    \headline = {%
        \headfont%
        \ifnum\pageno = 1%
            \hss%
        \else%
            \ifodd\pageno%
                \oddtitle\hss\folio%
            \else%
                \folio\hss\eventitle%
            \fi%
       \fi}%
    %\headline = {\hss}%
    \footline = {\hss}%
\fi%
\hfuzz     = 2pt%
\tolerance = 500%
\abovedisplayskip      = 3mm plus 6pt minus 4pt%
\belowdisplayskip      = 3mm plus 6pt minus 4pt%
\abovedisplayshortskip = 0mm plus 6pt%
\belowdisplayshortskip = 2mm plus 4pt minus 4pt%
\predisplaypenalty     = 0%
%
% ... symbol definitions:
%
%
\def\la{%
    \mathrel{\hbox{\rlap{\hbox{\lower4pt\hbox{$\sim$}}}\hbox{$<$}}}}%
\def\ga{%
    \mathrel{\hbox{\rlap{\hbox{\lower4pt\hbox{$\sim$}}}\hbox{$>$}}}}%
\def\utw{%
    \smash{\rlap{\lower5pt\hbox{$\sim$}}}}%
\def\udtw{%
    \smash{\rlap{\lower6pt\hbox{$\approx$}}}}%
\def\farcs{%
    \hbox{$.\!\!^{\prime\prime}$}}%
\def\getsto{%
    \mathrel{\hbox{\rlap{$\gets$}\hbox{\raise2pt\hbox{$\to$}}}}}%
\def\lid{%
    \mathrel{\hbox{\rlap{\hbox{\lower4pt\hbox{$=$}}}\hbox{$<$}}}}%
\def\gid{%
    \mathrel{\hbox{\rlap{\hbox{\lower4pt\hbox{$=$}}}\hbox{$>$}}}}%
\def\sol{%
    \mathrel{\hbox{\rlap{\hbox{\raise4pt\hbox{$\sim$}}}\hbox{$<$}}}}%
\def\sog{%
    \mathrel{\hbox{\rlap{\hbox{\raise4pt\hbox{$\sim$}}}\hbox{$>$}}}}%
\def\lse{%
    \mathrel{\hbox{\rlap{\hbox{\raise4pt\hbox{$<$}}}\hbox{$\simeq$}}}}%
\def\gse{%
    \mathrel{\hbox{\rlap{\hbox{\raise4pt\hbox{$>$}}}\hbox{$\simeq$}}}}%
\def\grole{%
    \mathrel{\hbox{\lower2pt\hbox{$<$}}\kern-8pt\hbox{\raise2pt\hbox{$>$}}}}%
\def\leogr{%
    \mathrel{\hbox{\lower2pt\hbox{$>$}}\kern-8pt\hbox{\raise2pt\hbox{$<$}}}}%
\def\loa{%
    \mathrel{\hbox{\rlap{\hbox{\lower4pt\hbox{$\approx$}}}\hbox{$<$}}}}%
\def\goa{%
    \mathrel{\hbox{\rlap{\hbox{\lower4pt\hbox{$\approx$}}}\hbox{$>$}}}}%
%
%
%
% ... font definitions:
% ... vector fonts:
%
\font\kleinhalbcurs=cmmib10 scaled 833%
%
% ... petit fonts:
%
\font\eightrm = cmr8%
\font\sixrm   = cmr6%
\font\eighti  = cmmi8%
\font\sixi    = cmmi6%
\font\eightsy = cmsy8%
\font\sixsy   = cmsy6%
\font\eightbf = cmbx8%
\font\sixbf   = cmbx6%
\font\eighttt = cmtt8%
\font\eightsl = cmsl8%
\font\eightit = cmti8%
\font\bxfr    = cmr10%
\font\bxft    = cmsl10%
\font\bxfm    = cmbx10%
\skewchar\eighti  = '177%
\skewchar\sixi    = '177%
\skewchar\eightsy = '60%
\skewchar\sixsy   = '60%
\hyphenchar\eighttt = -1%
%
% ... definition of versal greek letters:  
%
\mathchardef\Gamma   = "0100%
\mathchardef\Delta   = "0101%
\mathchardef\Theta   = "0102%
\mathchardef\Lambda  = "0103%
\mathchardef\Xi      = "0104%
\mathchardef\Pi      = "0105%
\mathchardef\Sigma   = "0106%
\mathchardef\Upsilon = "0107%
\mathchardef\Phi     = "0108%
\mathchardef\Psi     = "0109%
\mathchardef\Omega   = "010A%
%
% definition of figure environment:
%
\newif\ifwid%
\def\frame#1#2#3{%
    \vbox{\ifFrames\hrule height#2\fi%
          \hbox{\ifFrames\vrule width#2\fi%
                \vbox{\vskip#1 #3}%
          \ifFrames\vrule width#2\fi}%
    \ifFrames\hrule height#2\fi}}%
\def\begfigwid#1#2\endfig{%
    \ifReferee%
        \setbox2=\vbox{\box2\bigskip #2}%
    \else%
    \par\widtrue%
    \ifPostScript%
        \setbox1=\vbox{$$\frame{0pt}{0.5pt}{%
                       \epsfxsize=15truecm\epsffile{#1}}$$#2}%
    \else%
        \setbox1=\vbox{$$\frame{1cm}{0.5pt}{%
                       \vbox to 0pt{\hsize=141truemm\hfill}}$$#2}%
    \fi%
    \dimen0 = \ht1%
    \advance\dimen0 by \dp1%
    \advance\dimen0 by 5\baselineskip%
    \advance\dimen0 by 0.4 truecm%
    \ifdim\dimen0 > \vsize%
        \pageinsert\box1\vfill\endinsert%
    \else%
        \dimen0 = \pagetotal%
        \ifdim\dimen0 < \pagegoal%
            \advance\dimen0 by \ht1%
            \advance\dimen0 by \dp1%
            \advance\dimen0 by 1.4 truecm%
            \ifdim\dimen0 > \vsize%
                  \topinsert\box1\endinsert%
            \else%
                  \vskip 1 truecm\box1\vskip 4 truemm%
            \fi%
      \else%
            \vskip 1 truecm\box1\vskip 4 truemm%
      \fi%
\fi\fi}%
\let\begfig=\begfigwid%
%
% ... figure captions:
% ... (centered if text occupies less than one line)
%
\def\figure#1#2{%
    \ifenglish\def\fcap{Figure}\else\def\fcap{Abbildung}\fi%
    \ifReferee%
        \noindent{\fcap\ts#1.--\ }\ignorespaces #2%
    \else%
        \ifwid%
            \smallskip%
            \setbox0 = \vbox{%
                \noindent\petit{\fcap\ts#1.--\ }%
                \ignorespaces#2%
                \smallskip%
                \count255=0\global\advance\count255 by \prevgraf}%
            \ifnum\count255 > 1%
                \box0%
            \else%
                \centerline{%
                    \petit{\fcap\ts#1.--\ }\ignorespaces#2}%
                \smallskip%
            \fi%
        \else%
            \setbox0 = \vbox{%
                \hsize=75truemm\noindent\petit{\fcap\ts#1.--\ }%
                \ignorespaces #2%
                \smallskip%
                \count255=0\global\advance\count255 by \prevgraf}%
            \box0%
        \fi%
    \fi}%
%
% ... table environment:
% ... table captions:
%
\def\tabcap#1#2{%
    \ifenglish\def\tcap{Table}\else\def\tcap{Tabelle}\fi%
    \smallskip\vbox{%
        \noindent\petit{\tcap\ts#1\unskip.--\ }%
        \ignorespaces #2\medskip}}%
\def\begtabfull{%
    \par%
    \ifvoid%
        \topins\bgroup\midinsert\medskip%
    \else%
        \bgroup\topinsert\medskip%
    \fi}%
\def\endtab{%
    \endinsert\egroup\medskip\noindent}%
\def\begtabtest#1#2\endtab{%
    \par%
    \dimen0 = \hsize%
    \advance\dimen0 by -90 truemm%
    \advance\dimen0 by  -1 truecc%
    \relax\bgroup%
    \def\tabcap##1##2{%
        \vbox{\hsize=\dimen0\relax%
            \noindent\petit{\bf\tcap\ts##1\unskip.\ }%
            \ignorespaces##2\par}}%
    \topinsert\line{%
        \vbox{\hsize=90truemm\relax #1}%
        \hss#2\unskip}%
        \endinsert%
    \egroup\ignorespaces}%
%
% ... reference environment:
%
\def\begpet{%
    \vskip6pt\bgroup\petit}%
\def\endpet{%
    \vskip6pt\egroup}%
\def\begref#1{%
    \ifenglish\def\rcap{References}\else\def\rcap{Literatur}\fi%
    \sec{\rcap}\bgroup%
    \ifReferee\relax\else%
        \petit%
        \let\it  = \rm%
        \let\bf  = \rm%
        \let\sl  = \rm%
        \let\INS = N%
    \fi}%
\def\ref{%
    \par\noindent\hangindent=24.0pt\hangafter=1}%
\let\endref=\endpet%
%
% ...
%
\def\vec#1{%
    \hbox{%
        \textfont1   = \tamssm%
        \scriptfont1 = \kleinhalbcurs%
        \textfont0   = \bxfm%
        \scriptfont0 = \sevenbf%
        $#1$}}%
%
%
%
% ... smaller font for captions:
%
\def\petit{%
    \def\rm{\fam0\eightrm}%
    \def\it{\fam\itfam\eightit}%
    \def\sl{\fam\slfam\eightsl}%
    \def\bf{\fam\bffam\eightbf}%
    \def\tt{\fam\ttfam\eighttt}%
    \textfont0 = \eightrm%
    \textfont1 = \eighti%
    \textfont2 = \eightsy%
    \textfont\itfam = \eightit%
    \textfont\slfam = \eightsl%
    \textfont\bffam = \eightbf%
    \textfont\ttfam = \eighttt%
    \scriptfont0 = \sixrm%
    \scriptfont1 = \sixi%
    \scriptfont2 = \sixsy%
    \scriptfont\bffam = \sixbf%
    \scriptscriptfont0 = \fiverm%
    \scriptscriptfont1 = \fivei%
    \scriptscriptfont2 = \fivesy%
    \scriptscriptfont\bffam = \fivebf%
    \normalbaselineskip=9pt%
    \setbox\strutbox = \hbox{\vrule height7pt depth2pt width0pt}%
    \normalbaselines\rm%
    \def\vec##1{%
        \setbox0=hbox{$##1$}\hbox{\hbox to 0pt{%
            \copy0\hss}\kern0.45pt\box0}}}%
\let\ts=\thinspace%
%
% ... fonts for headlines:
%
\font\tafontr  = cmcsc10 scaled \magstep1%
\font\tafontt  = cmr10   scaled \magstep2%
\font\tafonts  = cmr7    scaled \magstep2%
\font\tafontss = cmr5    scaled \magstep2%
\font\tamt     = cmmi10  scaled \magstep2%
\font\tams     = cmmi10  scaled \magstep1%
\font\tamss    = cmmi10%
\font\tamssm   = cmmib10%
\font\tast     = cmsy10  scaled \magstep2%
\font\tass     = cmsy7   scaled \magstep2%
\font\tasss    = cmsy5   scaled \magstep2%
\font\tasyt    = cmex10  scaled \magstep2%
\font\tasys    = cmex10  scaled \magstep1%
\font\tbfontr  = cmcsc10 scaled \magstep1%
\font\tbfontt  = cmr10   scaled \magstep1%
\font\tbfonts  = cmr7    scaled \magstep1%
\font\tbfontss = cmr5    scaled \magstep1%
\font\tbst     = cmsy10  scaled \magstep1%
\font\tbss     = cmsy7   scaled \magstep1%
\font\tbsss    = cmsy5   scaled \magstep1%
\newbox\chsta%
\newbox\chstb%
\newbox\chstc%
\def\centerpar#1{%
    {\advance\hsize by-2\parindent%
        \rightskip = 0pt plus 4em%
        \leftskip  = 0pt plus 4em%
        \parindent = 0pt%
        \setbox\chsta = \vbox{#1}%
        \global\setbox\chstb = \vbox{%
            \unvbox\chsta\setbox\chstc=\lastbox%
            \line{\hfill\unhbox\chstc\unskip\unskip\unpenalty\hfill}}}%
    \leftline{\kern\parindent\box\chstb}}%
%
% ... headlines:
%
\def\chap#1{%
    \goodbreak%
    \ifReferee%
        \medskip%
    \else%
        \vskip24pt plus 6pt minus 4pt%
    \fi%
    \bgroup%
        \textfont0   = \tafontt%
        \scriptfont0 = \tafonts%
        \scriptscriptfont0=\tafontss%
        \textfont1=\tamt%
        \scriptfont1=\tams%
        \scriptscriptfont1=\tamss%
        \textfont2=\tast%
        \scriptfont2=\tass%
        \scriptscriptfont2=\tasss%
        \textfont3=\tasyt%
        \scriptfont3=\tasys%
        \scriptscriptfont3=\tenex%
        \baselineskip=18pt%
        \lineskip=18pt%
        \raggedright%
        \pretolerance=10000%
        \noindent%
        \ifReferee%
            \centerline{\tafontr\ignorespaces #1}%
        \else%
            \tafontt\ignorespaces #1%
        \fi%
        \vskip 7 truemm plus 6pt minus 4pt%
    \egroup%
    \noindent\ignorespaces}%
\def\sec#1{%
    \goodbreak%
    \vskip25 truept plus 4pt minus 4pt%
    \bgroup%
        \textfont0=\tbfontt %
        \scriptfont0=\tbfonts %
        \scriptscriptfont0=\tbfontss%
        \textfont1=\tams %
        \scriptfont1=\tamss %
        \scriptscriptfont1=\kleinhalbcurs%
        \textfont2=\tbst %
        \scriptfont2=\tbss %
        \scriptscriptfont2=\tbsss%
        \textfont3=\tasys %
        \scriptfont3=\tenex %
        \scriptscriptfont3=\tenex%
        \baselineskip=16pt%
        \lineskip=16pt%
        \raggedright%
        \pretolerance=10000%
        \noindent%
        \ifReferee%
            \centerline{\tbfontr\ignorespaces #1}%
        \else%
            \tbfontt\ignorespaces #1%
        \fi%
        \vskip 12 truept plus 4pt minus 4pt%
    \egroup%
    \noindent\ignorespaces}%
\def\subs#1{%
    \goodbreak%
    \vskip 15 truept plus 4pt minus 4pt%
    \bgroup%
        \noindent%
        \raggedright%
        \pretolerance=10000%
        \ifReferee\centerline{\bxfr\ignorespaces #1}%
        \else\bxft\ignorespaces #1\fi%
        \vskip 6 truept plus 4pt minus 4pt%
    \egroup%
    \noindent\ignorespaces}%
\def\subsubs#1{%
    \goodbreak%
    \vskip 15 truept plus 4pt minus 4pt%
    \bgroup%
        \sl%
        \noindent%
        \ignorespaces #1%
        \unskip.\ %
    \egroup%
    \ignorespaces}%
%
% ... footnote environment:
%
\def\footnoterule{%
    \kern-3pt%
    \hrule width 2 truecm%
    \kern2.6pt}%
\newcount\footcount%
\footcount = 0%
\def\advftncnt{%
    \advance\footcount by 1%
    \global\footcount = \footcount}%
\def\fonote#1{%
    \advftncnt%
    $^{\the\footcount}$%
    \begingroup%
        \petit%
        \def\textindent##1{%
            \hang\noindent\hbox to \parindent{##1\hss}\ignorespaces}%
        \vfootnote{$^{\the\footcount}$}{#1}%
    \endgroup}%
%
% ... acknowledgement:
%
%
%----------------------------------------------------------------------
% final message:
%----------------------------------------------------------------------
\newcount\sterne%
\outer\def\bye{%
    \bigskip\typeset%
    \sterne = 1%
    \ifx%
        \speciali\undefined%
    \else%
        \bigskip%
        Special characters created by the author%
        \loop\smallskip\noindent %
        Special character No\number\sterne:%
        \csname special\romannumeral\sterne\endcsname%
        \advance\sterne by 1%
        \global\sterne=\sterne%
        \ifnum\sterne < 11\repeat%
    \fi%
    \vfill\supereject\end}%
\def\typeset{%
    \centerline{\petit%
    This article was processed by the author using the \TeX\ %
    Macropackage aua.mac.}}%
%
% ... some A&A-like definitions for the title page:
%
\newif\ifauthor%
\def\MAINTITLE#1{\vfill\supereject\treset%
    \ifReferee\def\eventitle{}\else\def\eventitle{#1}\fi\chap{#1}}%
\def\AUTHOR#1{%
    \authortrue%
    \ifReferee\centerline{\tafontr #1}\else\noindent{\sl #1}\fi%
    \bigskip}%
\def\INSTITUTE#1{%
    \authorfalse%
    \noindent\hangindent=12pt\hangafter=1{#1}\bigskip}%
\def\ABSTRACT#1{%
    \ifenglish%
        \def\zcap{Abstract}%
    \else%
        \def\zcap{Zusammenfassung}%
    \fi%
    \ifReferee%
	\centerline{\tafontr\zcap}\medskip\noindent%
        \ignorespaces#1%
    \else%
        \hbox{\vbox{\hsize=0.2\hsize\hfill}\vbox{\hsize=0.8\hsize%
        \noindent{\sl\zcap.~}#1}}%
    \fi%
    \medskip}%
\def\KEYWORDS#1{%
    \noindent\ifReferee{\it Subject headings:~}\else{\sl Key words:~}\fi%
    {#1}\medskip}%
\def\at#1{%
    \ifauthor%
        $^{#1}$%
    \else%
        \par\noindent\hangindent=12pt\hangafter=1$^{#1}$%
    \fi}%
\def\maketitle{%
    \bigskip\hrule\bigskip}%
%
% ... redefinitions of headlines:
%
\newcount\anum\anum = 0        %
\newcount\bnum\bnum = 0         %
\newcount\cnum\cnum = 0      %
\newcount\rnum\rnum = 0%
\def\treset{\global\anum=0\global\footcount=0}%
\def\areset{\global\bnum=0}%
\def\breset{\global\cnum=0}%
\def\titlea#1{%
    \global\advance\anum by 1\areset%
    \ifReferee\def\oddtitle{}\else\def\oddtitle{#1}\fi%
    \sec{\the\anum.~#1}}%
\def\titleb#1{%
    \global\advance\bnum by 1\breset%
    \subs{\the\anum.\the\bnum.~#1}}%
\def\titlec#1{%
    \global\advance\cnum by 1%
    \subsubs{\the\anum.\the\bnum.\the\cnum.~#1}}%
\def\appendix#1{%
    \ifenglish%
       \def\xcap{Appendix}%
    \else%
       \def\xcap{Anhang}%
    \fi%
    \def\oddtitle{\xcap.~#1}\sec{\xcap.~#1}}%
\def\undertext#1{%
    \vtop{\hbox{#1}\kern 1pt \hrule}}%
\def\REF#1{%
    \global\advance\rnum by 1\ref{\the\rnum.~#1}}%
%
%
% ...
%
\def\begfigside#1#2\endfig{%
    \ifReferee\setbox2=\vbox{\box2\bigskip #2}\else%
    \par%
    \dimen0 = \hsize%
    \advance\dimen0 by -90 truemm%
    \advance\dimen0 by  -1 truecc%
    \relax%
    \bgroup%
        \def\figure##1##2{%
            \ifenglish%
                \def\fcap{Figure}%
            \else%
                \def\fcap{Abbildung}%
            \fi%
            \vbox{%
                \hsize = \dimen0\relax%
                \noindent\petit{\fcap\ts##1.--\ }%
                \ignorespaces ##2%
                \par}}%
        \midinsert\line{\vbox{%
            \hsize=90truemm\relax%
            \ifPostScript%
                \frame{0pt}{0.5pt}{%
                    \epsfxsize=\hsize\epsffile{#1}}%
            \else%
                \frame{1cm}{0.5pt}{%
                    \vbox to 0pt{\hsize=90truemm\hfill}}%
            \fi}%
            \hss#2\unskip}%
        \endinsert%
    \egroup\ignorespaces\fi}%
\def\begfigpage#1#2#3{%
    \pageinsert%
    \setbox0 = \vbox{%
        \ifenglish%
            \def\fcap{Fig}%
        \else%
            \def\fcap{Abb}%
        \fi%
        \smallskip\noindent\petit{\bf\fcap.\ts#2\ }%
        \ignorespaces #3}%
    \dimen0 = \vsize%
    \advance\dimen0 by -\ht0%
    \advance\dimen0 by -1cm%
    \ifPostScript%
        \setbox1 = \vbox{$$\frame{0pt}{0.5pt}{%
                       \epsfysize=\dimen0\epsffile{#1}}$$}%
    \else%
        \setbox1 = \vbox{$$\frame{0pt}{0.5pt}{\vbox to \dimen0{%
                       \hsize=141truemm\hfill}}$$}%
    \fi%
    \box1\vfill\box0\endinsert}%
\def\today{%
    \ifenglish%
        \number\day~%
        \ifcase\month%
            \or January  \or February\or March   \or April%
            \or May      \or June    \or July    \or August%
            \or September\or October \or November\or December%
        \fi%
        ~\number\year%
    \else%
        \number\day.~%
        \ifcase\month%
        \or Januar   \or Februar\or M\"arz  \or April%
        \or Mai      \or Juni   \or Juli    \or August%
        \or September\or Oktober\or November\or Dezember%
        \fi%
        ~\number\year%
    \fi}%
\MAINTITLE{Very close pairs of Quasi-Stellar Objects}
\AUTHOR{G. Burbidge\at{1}, F. Hoyle\at{2}, and P. Schneider\at{3}}
\INSTITUTE{\at{1}Center for Astrophysics and Space Sciences and
Department of Physics, University of California, La Jolla, California
92093-0111, U.S.A.; \at{2}102 Admirals Walk, Bournemouth BH2 5HF
Dorset, England; \at{3}Max Planck Institut f\"{u}r Astrophysik,
Karl-Schwarzschild-Strass 1, 85740 Garching bei M\"{u}nchen, Germany}
\ABSTRACT{It is pointed out that there are now known four very close
pairs of QSOs with separations $<5$ arcsec and very different
redshifts.  Several estimates of the probability that they are
accidental configurations range between $10^{-7}$ and $3.5 \times
10^{-3}$.  We conclude either that this is further evidence that QSOs
have significant non-cosmological redshift components, or that the
pairs must be explained by gravitational lensing.}
\titlea{Introduction}
If QSOs have redshifts entirely of cosmological origin and are randomly 
distributed in space, we shall expect to find very few very close pairs 
with very different redshifts.  The number depends on the surface density 
of QSOs, $\Gamma$, and the number of fields that have been examined (N), 
so that the number expected by accident $n$ is given by
$$
n = 2.42 \times 10^{-7} \Gamma \theta^{2} N ,
\eqno(1)$$
where $\theta$ is measured in arc seconds and $\Gamma$ is the number per 
square degree.

Thus when the first QSO pair 1548+115A,B was discovered (Wampler et al. 
1973), it was considered to be a strong argument in favor of 
non-cosmological QSO redshifts:  its two components have separation of 
$4\farcs8$, and their redshifts are 
$z_A = 0.44 \; {\rm and} \; z_B = 1.90$.  The probability to find 
such a close pair of QSOs among the $\sim 250$ QSOs then identified was 
estimated to be about 1\% if QSOs are distributed randomly on the sky.

In the following $\sim 20$ years the number of QSOs with measured 
redshifts has increased to more than 7000 (cf Hewitt \& Burbidge 1993).  
Also the gravitational lens phenomenon has been discovered and several 
close pairs with identical redshifts are known (see Keeton \& Kochanek 
1996 for a recent compilation of gravitationally lensed QSOs and 
candidate systems).  Added to this are a number of double QSOs with 
nearly identical redshifts  which are likely to be genuine QSO 
pairs and not lensed pairs since their spectra are not identically equal 
(cf Schneider 1994).  These pairs are usually attributed to the spatial 
two-point correlation between QSOs.

Comparatively recently three more very close pairs with very different 
redshifts have been discovered.
In Section 2 we describe and discuss them and look at the probability 
that they are accidental configurations.
In Section 3 we discuss all of the possible interpretations and 
implications of the results.
\titlea{The Observational Data and Probability Calculations}
Data on all four pairs of QSOs with very different redshifts are shown 
in Table 1.

\begtabfull
\tabcap{1}{Very Close Pairs of QSOs}
$$\vbox{\halign{
\hfil#\quad&\hfil#\quad&\hfil#\quad&\hfil#\quad&
\hfil#\quad&\hfil#\quad&\hfil#\cr
OBJECT & $m_{A}$ & $m_{B}$ & $z_{A}$ & $z_{B}$ & Separa- & Ref \cr
       &         &         &         &          & tions  & \cr
\noalign{\smallskip\hrule\smallskip}
0235$+$164A\&B & 14-19 & 19 & 0.94 & 0.52 
& $2\farcs5$ & (1) \cr
\hskip.2in & & & & &  & \cr
1009$-$025A\&B & 18.2 & 21.2 & 2.74 & 2.74 
& $1\farcs55$ & (2) \cr
1009$-$025A\&C & 18.2 & 19.3 & 2.74 & 1.62 
& $4\farcs6$ & (2)  \cr
\hskip.2in & & & & & &  \cr
\hskip.2in & & & & & &  \cr
1148$+$055A\&B & 17.9 & 20.7 & 1.89 & 1.41 
& $3\farcs9$ & (2) \cr
\hskip.2in & & & & &  & \cr
1548$+$114A\&B & 18.1 & 18.8 & 0.44 & 1.90 
& $4\farcs8$ & (3) \cr
\noalign{\smallskip\hrule}}}$$
\endtab
\vskip-1cm
\centerline{\petit (1) Burbidge et al. (1996), (2) Surdej et
al. (1994), (3) Wampler et al. (1973).}
\medskip
\ref {\bf AO 0235+164 A,B} This system was originally classified as a
BL Lac object with a second image often called a galaxy $2 \farcs 5$
away (Smith, Burbidge \& Junkkarinen, 1977; Cohen et al. 1987).  It
has recently been shown that the two components are a QSO (A) and QSO
or AGN (B) (Burbidge et al. 1996).  QSO A has long been known to be
rapidly varying at both radio and optical wavelengths, and A has two
optical absorption-line redshifts at $z = 0.524$ and 0.852.  The
absorption at $z = 0.524$ is also found in the 21 cm line and was
extensively studied by Wolfe, Davis \& Briggs (1982).  Several
candidate galaxies are close to it, one even closer than object B
(Stickel, Fried \& K\"{u}hr 1988, Yanny et al.  1989).  This object is
a strong continuum radio source.
\ref {\bf 1009--025 A,B,C} This system was discovered by Surdej et al.
(1994).  It has been entered in Table 1 as two pairs.  In the spectra
of 1009--025 A and B there are absorption redshifts at $z = 0.87 \;
{\rm and} \; z = 1.62$.  This pair then suggests an interpretation as
a gravitational lens.  However, the pair 1009--025 A and C or for that
matter the pairs 1009-025 B and C have very different redshifts and
the separation of A and C is only $4\farcs6$.
\ref {\bf 1148+055 A,B} This system was also discovered by Surdej et
al. (1994).
\ref {\bf 1548+115 A,B} As was previously mentioned this system was
discovered by Wampler et al. (1973).  It was one of a sample of 280 4C
radio sources in the identification program of Hazard et al. (1973).
There are a number of galaxies about $10^{\prime \prime}$ from
1548+114 A which have redshifts $z \simeq 0.434$ (Stockton 1974), very
close to the emission redshift of 1548+114 A.  The spectrum of
1548+114 B contains absorption at redshifts of 1.892, 1.756, 1.609 and
1.423 (Shaver \& Robertson 1985).
\par\noindent
Thus two of the four close pairs involve radio-emitting QSOs which
are very rare in comparison with radio-quiet QSOs.  It is 
usually assumed that only $\sim$ 1\% of QSOs are strong radio emitters.

Also in three of the four pairs there is, in addition to the very different 
emission redshifts, an absorption redshift which has the same value as 
one of the emission redshifts.  
In AO 
0235+164 an absorption redshift of 0.524 in A is almost identical with 
the emission redshift of B.  In 1009+025 A there is an absorption 
redshift at 1.62 which is the emission redshift of C, and in 1548+114 the 
emission redshift of A,  0.436, is almost identical with the galaxy 
redshifts of 0.434.
\medskip
\noindent{\bf Probability Calculations}
\smallskip
On the assumption that QSOs have cosmological redshifts and are randomly 
distributed we can use equation (1) to estimate $n$ for each pair.
Provided $n \ll 1$, then $n \approx p(1)$, the probability to find one 
QSO within $\theta$ in a sample of $N$ `primary' QSOs.  We discuss the 
four pairs in turn.

AO 0235+164 was originally described as a BL Lac object.  However the 
recent work has shown that AO 0235+164 A is a rapidly variable QSO with 
an emission redshift and AO 0235+164B is an adjacent QSO or AGN.  Thus 
the system should be removed from the BL Lac category.  The number of 
QSOs which are known to be rapidly variable is very small, so that we 
put N = 100.  Thus we find that the probability that one member of this 
sample has a second QSO closer than $2\farcs5$ 
and brighter than $m_{B} = 19$ is $n = 4.5 \times 10^{-4}$.  A much 
more conservative approach is to take all 515 sources from the 1-Jansky 
catalog (K\"{u}hr et al. 1981) as the parent population; then this 
probability increases to $n = 2.3 \times 10^{-3}$.

The two QSO pairs 1009--025 and 1148+055 were found in an optical survey 
for gravitational lenses by Surdej et al. (1994).  In recent years, 
there have been four such optical surveys performed, all of which took 
basically the same strategy:  to look for companions around 
high-luminosity QSOs, since for those the magnification bias should 
increase the observed fraction of lensed sources.  Kochanek (1993) lists 
the surveys and the number of QSOs in each of them;  there is a 
considerable overlap of targets among the four surveys.  The total 
number of QSOs imaged in these surveys is $N = 648$.  The expected 
number of pairs, where the second QSO is brighter than $m_{B} = 20.7$ 
and lies within $3\farcs9$ of the primary QSO, 
is $n = 0.12$.  Similarly, the expected number of QSOs within 
$4\farcs6$ of the primary QSOs brighter than 
$m_{B} = 19.3$ is $n = 0.017$.  Even a most conservative estimate 
yields very low probabilities:  The probability to find two (or more) 
QSO companions brighter than $m = 20.7$ (where we assume the surface 
density of QSOs to be about 50 per square degree) within $5^{\prime 
\prime}$ of the 648 high-luminosity QSOs in these lens surveys is 
$p (\geq 2) \approx 0.038$.

QSO 1548+114 was selected out of a sample of 280 radio sources from the 
4C catalog.  Not all these sources are QSOs, so that $N < 280$.  As 
reported in Hazard et al. (1973), only 53 of the 280 radio sources had a 
blue stellar object within the positional error box on the POSS.  Hence 
we take $N = 53$.  The fainter of the QSOs in this pair has $m_{B} = 
18.8$; the number density of QSOs up to this magnitude is estimated to 
be about $\Gamma = 3$ (e.g., Hartwick \& Schade 1990).  Hence the 
expectation value of the number of pairs with separation $\leq \theta$ 
in the sample investigated  by Hazard et al. (1973) is $n \approx 
8.9 \times 10^{-4}$.

We are aware of the fact that these probabilities have been calculated 
a posteriori and they should be interpreted with care.  Since they come 
from three independent samples the simplest method is to multiply the 
probabilities.  This gives a total probability of $\underline{8 \times 
10^{-8}}$ to find these four close pairs.

Alternatively we could combine the samples so that the total number in 
the sample is $N < 1000$.  If we then put $\theta = 5^{\prime \prime}$ 
and $\Gamma = 50$ (corresponding to close companions brighter than 
$20.\!\!^{\rm m}7$, then $n = 0.3$ as compared with the four 
pairs which are found, the probability of which is $\sim n^{4}/4! = 2.7
\times 10^{-4}$. 

Most conservatively -- and one of the authors (PS) views this as the 
most legitimate combination of probabilities -- one might assume that a 
total of $N = 2000$ QSOs have been investigated for a close companion 
QSO with magnitude brighter than $m = 20.7$ (companions as faint as that 
will not be readily identified on the POSS!); then the probability of 
finding four (or more) companions within $5^{\prime \prime}$ of the 
primary QSOs is
$$
p (\geq 4) = 3.5 \times 10^{-3} \; ,
\eqno(2)$$
and the expected number of pairs is $n = 0.61$.

In the following section we consider ways of explaining the existence of 
these pairs.
\titlea{Possible Interpretations}
There are in principle three possible explanations for these phenomena.

\item{1.} In the framework of standard cosmology an enhancement of the 
number of close pairs with discordant redshifts can be obtained if the 
two-point correlation function extends over distances corresponding to 
the redshift differences.  However, the redshift differences in Table 
1 are so large that none of the presently discussed cosmologies would 
predict any appreciable correlations in these cases.
\item{2.} The results taken at their face value 
indicate that significant parts of the redshifts have 
a non-cosmological origin (cf. for example Burbidge 1996) and the pairs 
are physically associated.
\item{3.} Back to the cosmological interpretation, it must be argued that a 
local enhancement of the QSO density in some part of the sky can be 
caused by gravitational lensing which affects the apparent magnitude of 
QSOs and can lead to the preferential inclusion of lensed QSOs into 
flux-limited samples.

Since (1) is clearly ruled out, we are left with (2) and (3).  The 
authors of this paper have divergent views about the likelihood that (2) 
or (3) is the explanation.  Much evidence for the existence of 
non-cosmological redshifts has been discussed elsewhere (Hoyle \& 
Burbidge 1996; Burbidge 1996).

Thus we turn to (3) and discuss what can be said in favor of a 
gravitational lensing scenario.
\titlea{A Gravitational Lens Origin for Close QSO Pairs}
Gravitational light deflection can not only lead to the occurrence of 
multiply imaged QSO and radio galaxies, but it also affects the apparent 
magnitude of sources when there is a matter concentration in or near the 
line-of-sight to them.  An over-density of matter in the foreground of a 
source will magnify it.  Depending on the steepness of the source 
counts, this magnification can yield a dramatic biasing effect:  Sources 
which without lensing would be too faint to be included in a 
flux-limited sample can be boosted above the flux threshold and thus be 
included in the sample.  That is, magnified sources are preferentially 
included in flux-limited samples.  If the source counts are steep, then 
for every bright source there is a large number of faint sources, from 
which the magnified sources can be drawn.  Hence, this magnification 
bias is strong for steep counts, and unimportant for flat counts (for a 
detailed discussion and references on the magnification bias, see Sect. 
12.5 of Schneider, Ehlers \& Falco 1992).

It can be argued that at least two of the QSO pairs show strong evidence 
for lensing to be important.  This is most obvious in the QSO 1009--025, 
where the QSO with the larger redshift is multiply imaged.  In the 
spectra of the two QSO images, absorption lines are seen at redshift 
$z_{a} = 0.87$ and at $z_{a} = 1.62$ i.e., the redshift of the 
lower-redshift QSO (Hewett et al. 1994).  While the available 
information about this lens system is not sufficient for constructing a 
detailed lens model, it is likely that the higher-$z$ QSO is magnified 
by at least 1 mag, as is typical for double images.  In AO 0235+164, 
gravitational lensing has long been suspected, for example to account 
for the strong variability in the optical and the radio flux, which 
might find an explanation in terms of microlensing.  The long-known 
companion about $2^{\prime \prime}$ to the south of AO 0235+164A, several 
candidate galaxies even closer to it (Stickel, Fried \& K\"{u}hr 
1988, Yanny et al. 1989), and the observed 21 cm line absorption 
(Wolfe, Davis \& Briggs 1982) may be indications of potential lenses in 
this system; in fact, from the image of a galaxy only $\sim 
0\farcs5$ away from the BL Lac (Stickel et al. 
1988), one may ask why no multiple images are seen in this 
system (Narayan \& Schneider 1990).  Also, 
Iovino \& Shaver (1986) have placed 
upper bounds on the mass of the foreground QSO in the system 1548+114 
from the absence of a secondary image of the higher redshift QSO.

One can think of two variants of a lensing scenario:  in the first, the 
lenses are positioned at redshifts lower than both QSOs, i.e., both QSOs 
are magnified, and in the second, the lens is physically associated with 
the foreground QSO and magnifying only the background QSO.  From the 
preceding remarks about magnification bias, the former scenario is 
considered unlikely:  in three of the four pairs, the foreground QSO is 
at $m = 19$ or fainter, i.e., close to or beyond the break in the QSO 
number counts.  At these magnitudes, the magnification bias is very weak 
and can even lead to a decrease of the local number counts.  Hence, in 
the first scenario one would not expect to obtain an increased number of 
pairs from lensing.

A toy model should illustrate the possible effects of the second 
scenario:  Consider a `foreground sky' and a `background sky';  on the 
latter, the higher-redshift QSOs are randomly distributed, having 
unlensed source counts of the form $n ( > S ) \propto S ^{- \alpha}$, 
with $\alpha \approx 2.6$ (e.g., Hartwick \& Schade 1990).  Suppose that 
a fraction $f$ of the `foreground sky' contains matter over-densities 
which magnify QSOs on the `background sky' by a factor $\mu_{+}$, 
whereas in the other directions, background sources are (de)magnified by 
a factor $\mu_{-}$.  Flux conservation (Schneider et al. 1992, Sect. 
4.5.1) then requires that $\mu_{-} = \mu_{+}(1-f)/(\mu_{+} - f)$. 
Futhermore, assume that QSOs in the `foreground' are concentrated 
towards those directions in which over-densities of matter are present.  
That is, if $\bar{n}$ is the mean number density for  foreground QSOs, let 
the number density in the magnifying fraction of the `foreground sky' be 
$\nu_{+} \bar{n}$, whereas the number density in the rest of the sky is 
$\nu_{-} \bar{n} = (1-\nu_{+} f)/(1 - f)$, with $\nu_{+} \leq 1/f$.  Using 
the preceding assumptions, one can then show that in a flux-limited 
sample of $N$ background QSOs the expected number of foreground QSOs 
within an angle $\theta$ is
$$
n_{12}  =  Q \pi \theta^{2} N \bar{n}
\eqno(3)$$
where the factor
$$
Q  =  {f \nu_{+} (\mu _{+} - f) ^{\alpha - 1} + (1 - f) ^{\alpha 
-1} (1 - \nu_{+} f)\over f ( \mu_{+} - f)^{\alpha -1} + (1 - f)^{\alpha}}
\eqno(4)$$
describes the ratio of expected pairs relative to the case 
that no lensing takes place.  In Fig. 1, we have plotted $Q$ as a 
function of $f$, for the maximum value of $\nu_{+} = 1/f$, i.e. all QSOs 
in the foreground sky are assumed to lie in the over-dense regions.

\begfig{fig1.tps}
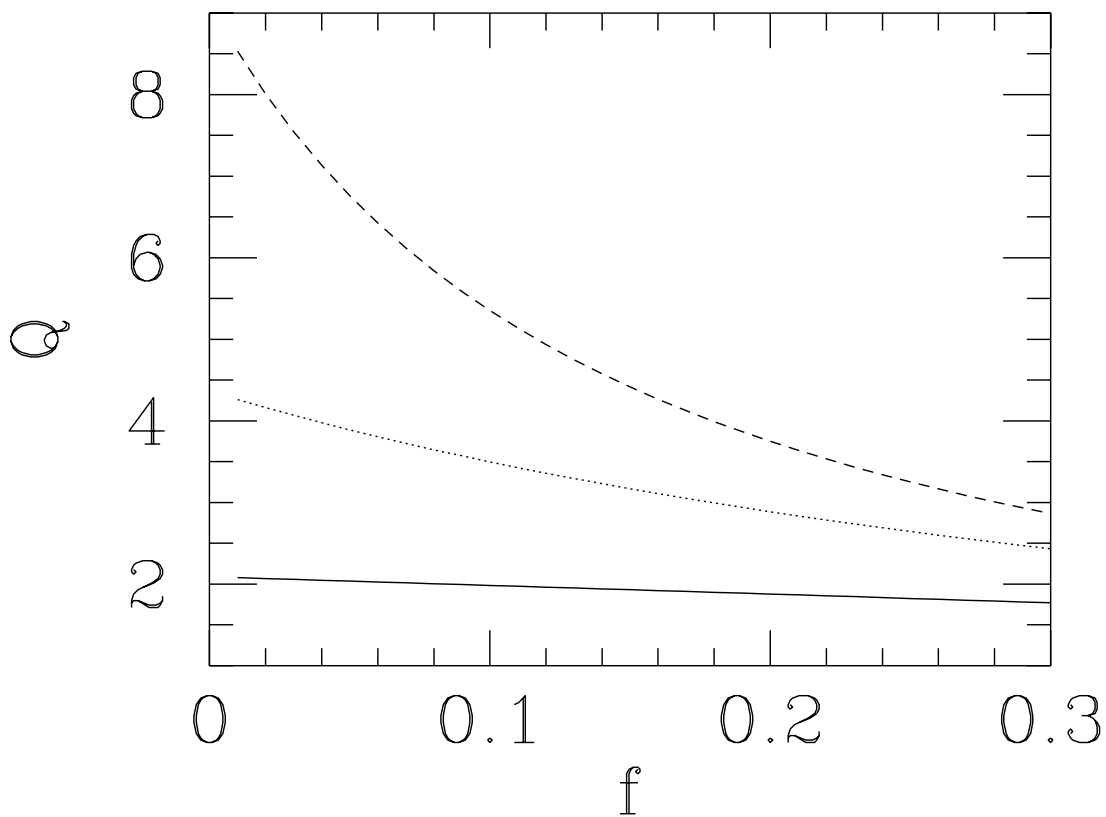
\figure{1}{The ratio $Q$ of pairs of foreground-background QSOs in
the lensing toy-model described in the text, relative to the case of
no lensing present, as a function of the fraction of the sky $f$ in
which overdensities of matter leads to magnification of the background
QSOs by a factor $\mu_+$. The solid (dotted, dashed) curve corresponds
to magnification of half a magnitude (one magnitude, 1.5 magnitudes),
and it has been assumed that all foreground QSOs are situated in the
overdense regions, $\nu_+=1/f$}
\endfig

As can be inferred from the figure, the increase in the expected number 
of pairs is quite substantial, even for low values of the magnification. 
For example, if the magnification in $f = 10$\% of the sky is one 
magnitude $( \mu_{+} = 10^{0.4})$, the expected number of pairs 
increases by a factor of about 3.5.  Such an increase would suffice to 
increase the probability in Eq. (2) to about 18\%, and hence the 
observed number of pairs would not pose an improbable statistical 
fluctuation.   It should be clear that the toy model presented here is 
not realistic, but it illustrates the basic features of a more realistic 
lensing scenario.  One of the basic problems encountered in making a 
realistic model is that the observed number density of QSOs flattens as 
we go to fainter magnitudes so that while $\alpha \simeq 2.6$ 
up to $m_{B} = 19.5$, it becomes $\alpha \leq 1$ for the range 19.5 
to 21.5 (Hartwick and Schade 1990).
\titlea{Conclusion}
We have shown that if the redshifts of the QSOs are of cosmological 
origin and gravitational lensing is not a factor, it is extremely 
improbable that the pairs could have these configurations by accident.  
If they are physically associated, and the lower emission redshift in 
each pair gives the true distance of the pair, then the intrinsic 
redshifts ($z_{i}$) of the higher redshift objects are:  $z_{i} = 0.27$ 
for AO 0235+164; $z_{i} = 0.43$ for 1009--025; $z_{i} = 0.19$ for 1148+055, 
and $z_{i} = 1.02$ for 1548+115.

Two of us (GB and FH) consider that the existence of these pairs is 
further strong evidence in favor of the view that QSOs often have 
redshift components of intrinsic origin.  One of us (PS) considers that 
while no realistic  model has yet been constructed it may still be 
possible to interpret these phenomena in terms of gravitational lensing 
of QSOs with cosmological redshifts.

One of us (GB) is grateful for hospitality afforded him at Max Planck 
Institut f\"{u}r Extraterrestrische Physik in September 1995. This
work was partially supported by the Sonderforschungsbereich SFB 375-95
of the Deutsche Froschungsgemeinschaft (PS).

\begref{References}
\ref Burbidge, G. 1996, A\&A, in press.
\ref  Burbidge, E.M., Beaver, E., Cohen, R.D., Hamann, F., 
Junkkarinen, V.T., Lyons, R., Zuo, L. 1996, B.A.A.S., in press.
\ref  Cohen, R., Smith, H.E., Junkkarinen, V. and Burbidge, E.M. 
1987, ApJ, 318, 577.
\ref  Hartwick, F.D.A. \& Schade, D. 1990, ARA\&A, 28, 437.
\ref  Hazard, C., Jauncey, D.L., Sargent, W.L.W., Baldwin, J.A. \& 
Wampler, E.J. 1973, Nat, 246, 205.
\ref  Hewett, P.C. et al. 1994, AJ, 108, 1534.
\ref  Hewitt, A. \& Burbidge, G. 1993, ApJS, 87, 451
\ref  Hoyle, F. \& Burbidge, G. 1996, A\&A, in press.
\ref  Iovino, A. \& Shaver, P. 1986, A\&A, 166, 119.
\ref  Keeton, C.R. \& Kochanek, C.S. 1996, in: {\it Astrophysical 
applications of gravitational lensing,} C.S. Kochanek \& J.N. Hewitt 
(eds), Kluwer:  Dordrecht, p. 419.
\ref  K\"{u}hr, H., Witzel, A., Pauliny-Toth, I.I.K. \& Nauber, U. 
1981, A\&AS, 45, 367.
\ref  Narayan, R. \& Schneider, P. 1990, MNRAS, 243, 192.
\ref  Schneider, P. 1994, in: {\it Gravitational Lenses in the 
Universe,} J. Surdej, D. Fraipont-Caro, E. Gosset, S., Refsdal \& M. 
Remy (eds), Proceedings of the 31st Li\`{e}ge International Astrophysical 
Colloquium, Li\`{e}ge, p. 41.
\ref  Schneider, P., Ehlers, J. \& Falco, E.E. 1992, {\it 
Gravitational Lenses}, Springer:  New York.
\ref  Shaver, P. and Robertson, J.G. 1985, MNRAS, 212, 15P.
\ref  Smith, H.E., Burbidge, E.M. \& Junkkarinen, V.T. 1977, ApJ, 218, 
611.
\ref  Stickel, M., Fried, J.W. \& K\"{u}hr, H. 1988, A\&A, 198, L13.
\ref  Stockton, A. 1974, Nature, 250, 308.
\ref  Surdej, J. et al. 1993, AJ, 105, 2064.
\ref  Surdej, J. et al. 1994, in: {\it Gravitational Lenses in the 
Universe}, J. Surdej, D., Fraipont-Caro, E. Gosset, S. Refsdal \& M. 
Remy (eds), Proceedings of the 31st Li\`{e}ge International 
Astrophysical Colloquium, Li\`{e}ge, p. 153.
\ref  Wampler, E.J., Baldwin, J.A., Burke, W.l., Robinson, L.B. \& 
Hazard, C. 1973, Nat 246, 203.
\ref  Wolfe, A.M., Davis, M.M. \& Briggs, F.H. 1982, ApJ, 259, 495.
\ref  Yanny, B., York, D.G. \& Gallagher, J.S. 1989, ApJ., 338, 735.
\endref
\end